# Docking Studies on HIV Integrase Inhibitors Based On Potential Ligand Binding Sites

Subhomoi Borkotoky*

Centre for bioinformatics, Pondicherry University, Puducherry, India

subhomoi@mails.bicpu.edu.in

## Abstract

*HIV integrase is a 32 kDa protein produced from the C-terminal portion of the Pol gene product, and is an attractive target for new anti-HIV drugs. Integrase is an enzyme produced by a retrovirus (such as HIV) that enables its genetic material to be integrated into the DNA of the infected cell. Raltegravir and Elvitegravir are two important drugs against integrase.*

## 1. Introduction

The human immunodeficiency virus type 1 (HIV-1) is the primary cause of the acquired immunodeficiency syndrome (AIDS), which is a slow, progressive and degenerative disease of the human immune system. HIV-1 is a lentivirus belonging to the retrovirus family. The virus is diploid and contains two plus-stranded RNA copies of its genome. The development of possible methods that can delay progression of the infection or block replication of HIV-1 in infected individuals has been the subject of dedicated research efforts over the past decades. One important issue is that HIV-1 makes use of the replication machinery of the host cell, which minimizes the number of potential viral targets. On the other hand, the close host-virus relationship limits the evolutionary freedom for the viral components that interact with the host molecules[1]. Integration of viral DNA into the host chromosome is a necessary process in the HIV replication cycle [2]. The key steps of DNA integration are carried out by the viral integrase protein, which, along with protease and reverse transcriptase, is one of three enzymes encoded by HIV. Combination antiviral therapy with protease and reverse transcriptase inhibitors has demonstrated the potential therapeutic efficacy of antiviral therapy for treatment for AIDS [3] . Since HIV integrase has no direct cellular counterpart it presents itself as an attractive target for therapeutic intervention [4]. Unlike the retroviral reverse transcriptase and protease enzymes, successful drug candidates based on the inhibition of integrase have yet to emerge despite the multitude of laboratories working on the problem [5]. The objective of this study is to discover new analogs with improved potency, physiochemical/metabolic properties and toxic effects, which can stand as potential inhibitors of AIDS.

## 2. Materials and Methods

For the present study bioinformatics tools, biological databases like PubMed, Drug Bank, PDB (Protein Data Bank) and software's like Molegro Virtual Docker, ACD ChemSketch, Pharma Algorithms were used.

The structure of HIV-1 integrase catalytic domain was retrieved from Protein data bank (PDB) (PDB id: 1BL3) (http://www.pdb.org). It is a 160 residues long protein and contains three domains, an N-terminal HH-CC zinc finger domain believed to be partially responsible for multimerization, a central catalytic domain and a C-terminal domain. Both the Central catalytic domain and C-terminal domains have been shown to bind both viral and cellular DNA. Currently no crystal structure data exists with Integrase bound to its DNA substrates. Biochemical data and structural data suggest that integrase functions as a dimer or a tetramer [6].

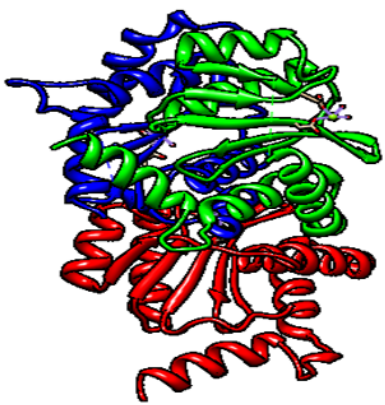

Figure 1. The structure of HIV-1 integrase catalytic domain

The structures of the drugs Raltegravir and Elvitegravir were obtained from Drug Bank (www.drugbank.ca) and KEGG Drug (www.genome.jp/kegg/drug) respectively. Using ACDLABS/ChemSketch (www.acdlabs.com) the 2D structures of the analogues of these drugs were sketched. The analogs were screened for their physiochemical properties at FAF drugs. The analogs that do not follow Lipinski's rules are discarded. The selected analogs are then screened for their bioactivity and toxicity under PASS software. The selected analogs were then searched against various chemical structure databases for similarity with a existing structure. The databases taken in this step are: PubChem (pubchem.ncbi.nlm.nih.gov), KEGG, Molsoft (MolCart) (www.molsoft.com/molcart.html), Hic-Up (xray.bmc.uu.se/hicup/), and ChemBank (chembank.broadinstitute.org/). No similarity has been detected. The analogs are then subjected to molecular dynamics analysis in ChemBio3D Ultra under optimum conditions. The screened analogs were then subjected to Pharma Algorithm predict various pharmacological effects like oral bioavailability, protein binding etc. The Molecular Docking is performed in Molegro Virtual Docker (MVD) (http://www.molegro.com/). Possible active sites and cavities were detected for Chain A of 1BL3 using Molegro Virtual Docker. The following Parameters were used for Cavity Detection:

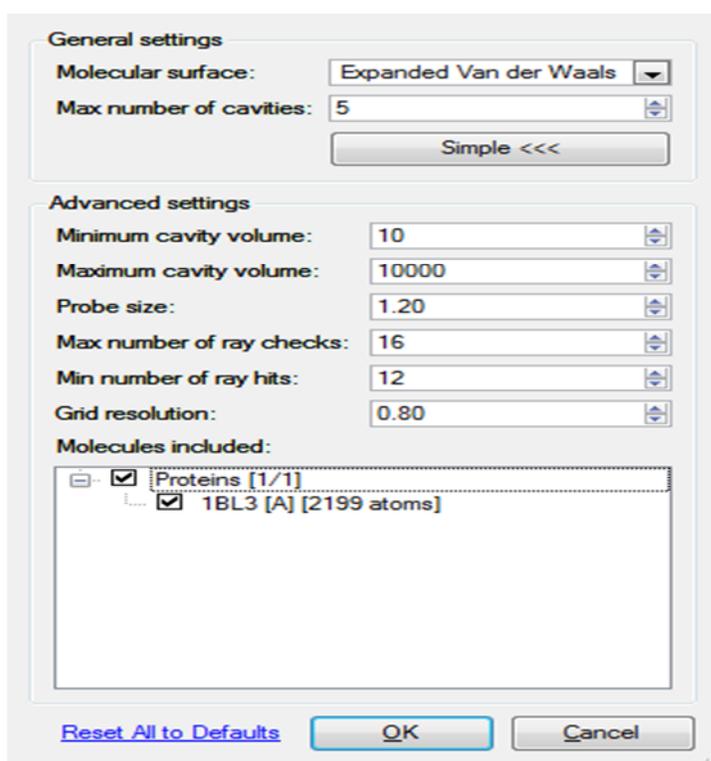

Figure 2. Parameters of Molegro Virtual Docker.

## 3. Result

### 3.1 Result of Active Site Prediction and Cavity Detection

The structure of HIV-1 integrase catalytic domain (PDB id: 1BL3) is 160 residues long and contains three chains: A, B and C. Catalytic core domain is present between 50 – 212 residues [7] . The position of the active site are Thr (T) =66. Asp (D) =64,116. Val (V) =77. Glu (E) =15 Lys (K) =159 [8, 9].  Active sites and cavities were detected for Chain A of 1BL3 using Molegro Virtual Docker.

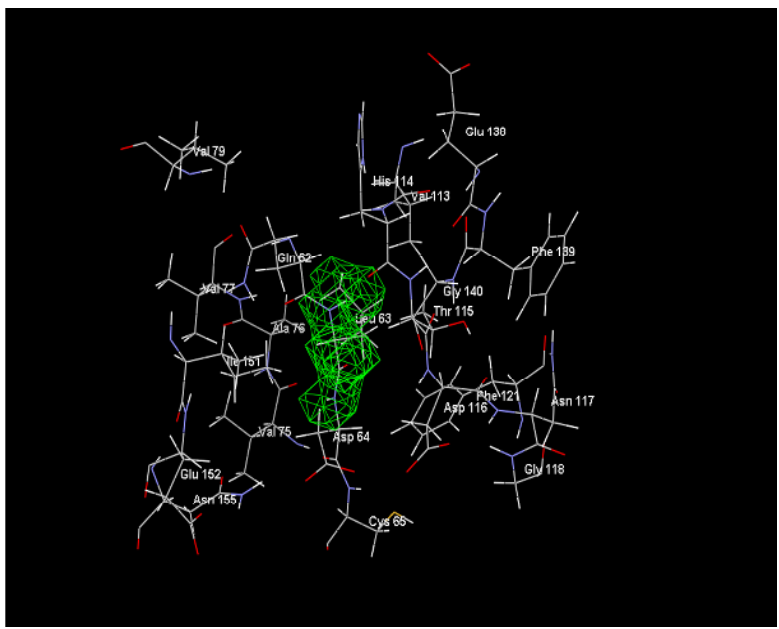

Figure 3. Positions of active site shown in Molegro Virtual Docker

## 3.2 Result of Molecular Docking:

The Molecular Docking is performed in Molegro Virtual Docker (MVD) .The following Parameters were used for Docking using Molegro Virtual Docker. Docker uses the MolDock docking engine to predict ligand - protein interactions. MolDock is based on a new hybrid search algorithm, called guided differential evolution [10].

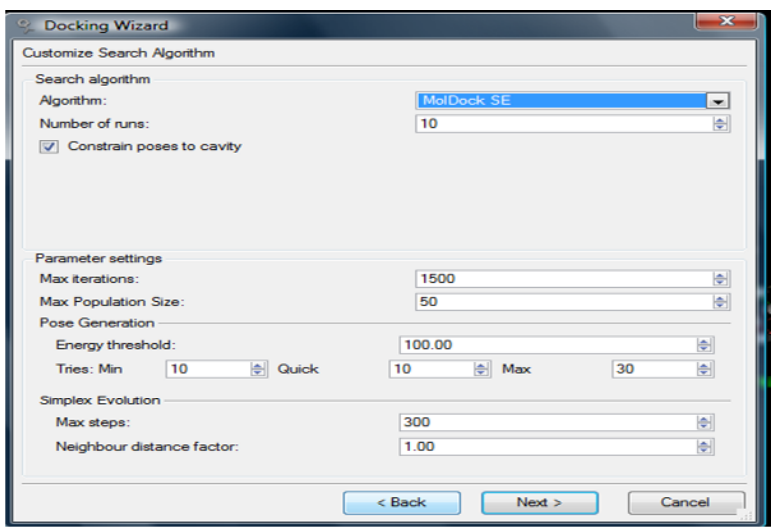

Figure 4. Docking Parameters shown in Molegro Virtual Docker

The imported ligands were manually checked before docking and corrected in those cases were it had failed. Water molecules with the protein structures were excluded from the docking experiments, The docking is then allowed to run for some time for Raltegravir and Elvitegravir analogs (remained after virtual screening) respectively with the target protein. The poses having the good MolDock and Docking score are selected for the further analysis.

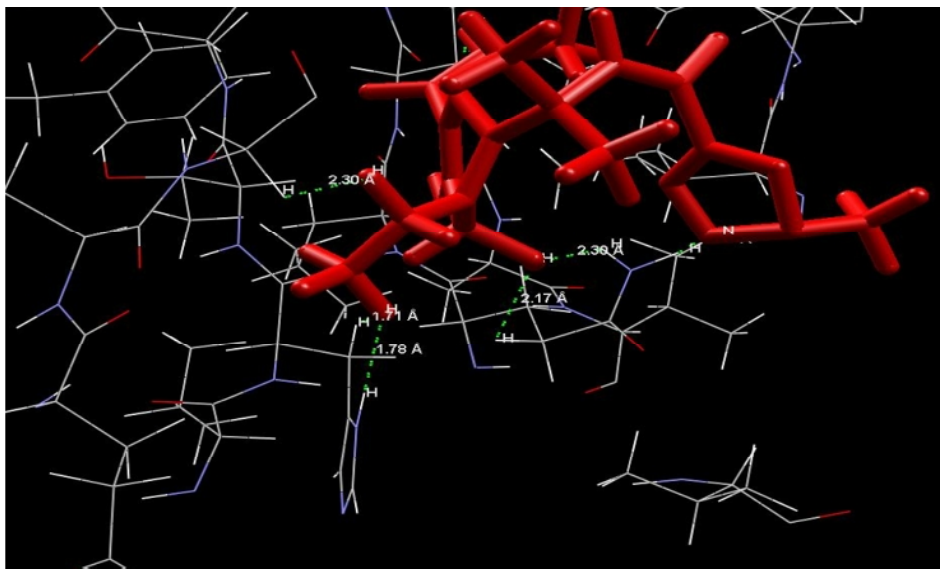

Figure 4**.** The result of Docking shown with contacts.

Docking results tabulated between HIV integrase and the conventional drug Raltegravir (Table 1) and Elvitegravir (Table 2) analogs are shown below.

Table 1. Raltegravir analogs

| Mol_id | MolDock Score | Docking score |
|---|---|---|
| RL 40 | -150.584 | -160.781 |
| RL 13 | -144.854 | -144.94 |
| RL 4 | -143.524 | -154.891 |
| RL 3 | -142.581 | -148.704 |

Table 2. Elvitegravir analogs

| Mol_id | MolDock Score | Docking score |
|---|---|---|
| EL 9 | -114.166 | -199.066 |
| EL 15 | -113.443 | -118.713 |
| EL 10 | -112.471 | -112.235 |
| EL 7 | -111.277 | 113.903 |

The interactions between the residues of the target protein and analogs showing good docking score which are derived from the two lead molecules are given below-

25

Table 3. Interaction of Raltegravir analogs

| Mol_id | Residue | Interaction | Distance (in Armstrong) | Mol_id | Residue | Interaction | Distance (in Armstrong) |
|---|---|---|---|---|---|---|---|
| RL 9 | Glu (152) | H - H | 1.57 | RL 28 | Glu (152) | H - H | 1.82 |
| | Glu (152) | H – H | 1.36 | | Glu (152) | H - H | 2.01 |
| | Gly (140) | H – O | 2.22 | | Glu (152) | H - H | 1.91 |
| | Asp(116) | H - O | 2.28 | | Asp(64) | H - H | 2.32 |
| RL 10 | Asp(116) | O – H | 2.19 | RL 40 | His(114) | H - H | 1.71 |
| | Asp(64) | H – H | 0.90 | | Gln(62) | N - H | 1.80 |
| | Gly(140) | H – O | 2.00 | | Asp(64) | H - H | 1.92 |
| | Gly(140) | H – H | 1.66 | | His(114) | H - H | 1.78 |

Table 4. Interaction of Elvitegravir analogs

| Mol_id | Residue | Interaction | Distance (in Armstrong) | Mol_id | Residue | Interaction | Distance (in Armstrong) |
|---|---|---|---|---|---|---|---|
| EL 7 | Glu (152) | H - H | 2.09 | EL 10 | His(114) | H - Cl | 2.38 |
| | Glu (152) | H - H | 2.36 | | Glu (152) | H - H | 2.25 |
| EL 9 | Glu (152) | H - H | 2.02 | EL 15 | His(114) | H - H | 1.57 |
| | Thr (115) | H - H | 2.08 | | Asp(116) | H - H | 1.87 |
| | Asp(64) | H - O | 2.32 | | Asp(64) | H - H | 2.36 |
| | Asp(64) | H - O | 2.37 | | His(114) | H - O | 2.15 |

### 3.4 Result of Pharma algorithm

Pharma algorithm is a tool for *in silico* physicochemical, ADME, Metabolism, and Toxicology screening and prediction. The results for Raltegravir (Table 5) and Elvitegravir (Table 6) analogs are shown below

Table 5. Pharma algorithm results for Raltegravir

| ADME BOX | Raltegravir | RL 40 | RL 28 | RL 10 | RL 9 |
|---|---|---|---|---|---|
| Oral bioavailability between 30% and 70% | | | | | |
| %F(Oral) > 30%: | 0.721 | 0.827 | 0.721 | 0.099 | 0.391 |
| %F(Oral) >70%: | 0.351 | 0.282 | 0.282 | 0.080 | 0.091 |
| Absorption | | | | | |

| Maximum passive absorption Contribution from: | | | | | |
|---|---|---|---|---|---|
| Trancellular route | 100% | 100% | 100% | 100% | 100% |
| Paracellular route | 0% | 0% | 0% | 0% | 0% |
| Permeability: | | | | | |
| Human Jejunum scale (pH=6.5) Pe, Jejunum = | $2.37 \times 10^{-4}$ cm/s | $2.62 \times 10^{-4}$ cm/s | $3.60 \times 10^{-4}$ cm/s | $2.35 \times 10^{-4}$ cm/s | $1.71 \times 10^{-4}$ cm/s |
| Caco-2 scale (pH=7.4, 500 rpm): Pe, Caco-2 | $61.04 \times 10^{-6}$ cm/s | $121.86 \times 10^{-6}$ cm/s | $159.45 \times 10^{-6}$ cm/s | $59.88 \times 10^{-6}$ cm/s | $27.88 \times 10^{-6}$ cm/s |
| Absorption rate Ka =: | 0.080 min-1 | 0.085 min-1 | 0.095 min-1 | 0.080 min-1 | 0.062 min-1 |

Table 6. Pharma algorithm results for Elvitegravir

| ADME BOX | Elvitegravir | EL27 | EL47 | EL28 | EL7 |
|---|---|---|---|---|---|
| Oral bioavailability between 30% and 70% | | | | | |
| %F(Oral) > 30%: | 0.888 | 0.827 | 0.888 | 0.541 | 0.541 |
| %F(Oral) >70%: | 0.480 | 0.388 | 0.820 | 0.121 | 0.110 |
| Absorption | | | | | |
| Maximum passive absorption Contribution from | | | | | |
| Trancellular route | 100% | 100% | 100% | 100% | 100% |
| Paracellular route | 0% | 0% | 0% | 0% | 0% |
| Permeability: | | | | | |
| Human Jejunum scale (pH=6.5) Pe, Jejunum = | $3.34 \times 10$-4 cm/s | $3.38 \times 10$-4 cm/s | $2.71 \times 10$ cm/s | $2.71 \times 10$ | $1.06 \times 10$ |
| Caco-2 scale (pH=7.4, 500 rpm): Pe, Caco-2 | $60.34 \times 10$-6 cm/s | $62.71 \times 10$-6 cm/s | $32.21 \times 10$ cm/s | $32.45 \times 10$ | $3.09 \times 10$ |
| Absorption rate Ka =: | 0.093 min-1 | 0.093 min-1 | 0.086 min-1 | 0.086 min$^{-1}$ | 0.033 min$^{-1}$ |

International Journal on Bioinformatics & Biosciences (IJBB) Vol.2, No.3, September 2012

The structures of the two drugs Raltegravir and Elvitegravir and the changes or modifications in the two analogs (RL-10, EL -7) showing good protein binding are shown below.

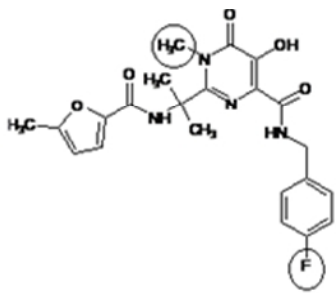
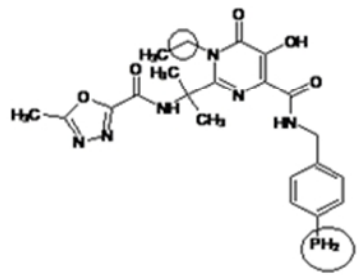

Figure 5.  The structure of Raltegravir          Figure 6. Structural analog of Raltegravir RL10

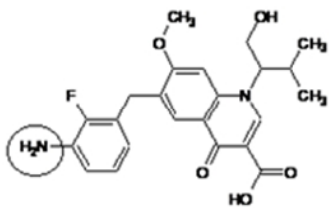
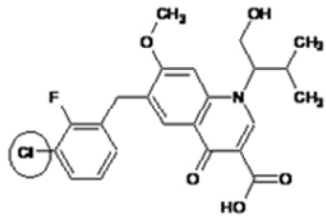

Figure 7.  The structure of Elvitegravir         Figure 8. Structural analog of Elvitegravir EL7

## Conclusion

From the molecular docking result carrying out using Molegro virtual docker, the best  four analog form each inhibitor ( Raltegravir- RL-40,RL-13, RL -10, RL-9 ; Elvitegravir- EL-27,EL-47,EL-28, EL-21 ) based on the mol dock score and the docking score  were selected. The close contacts shows that there is a high possibility of interaction of these analogs with the amino acids of the active site of the  protein .These analogs also show better bioavailability (RL 40 ; EL47) , protein binding  (RL 10, EL 7) ,solubility  (EL7, RL 10 ) and low toxicity than existing two inhibitors  ( Raltegravir , Elvitegravir). These inhibitor analogs will provide a platform for structure-based design of an additional class of inhibitors for antiviral therapy.